\documentclass[12pt]{article}

\usepackage{amssymb}
\usepackage{amsmath}

\usepackage{epsfig}

\begin{document} %

\title{\bf Randall-Sundrum scenario with small curvature and dilepton
 production at LHC}

\author{A.V. Kisselev\thanks{\ Talk presented at the International School-Seminar ``New Physics
and Quantum Chromodynamics at External Conditions'', May 22-24,
2013, Dnipropetrovsk, Ukraine.
} \\
{\small Institute for High Energy Physics, 142281 Protvino, Russia}
\\
{\small and}
\\
{\small Department of Physics, Moscow State University, 119991
Moscow, Russia}}

\date{}

\maketitle

\thispagestyle{empty}


\begin{abstract}
The brief review of the recent results obtained in the
Randall-Sundrum scenario with the small curvature of the
five-dimensional space-time is presented.
\end{abstract}


\section{RSSC model}

In the recent papers \cite{Kisselev:Kisselev:dimuon},
\cite{Kisselev:Kisselev:dielectron} the dilepton (dimuon and
dielectron) production at the LHC was studied in the framework of
the \emph{RSSC model} (Randall-Sundrum-like model with the small
curvature) \cite{Kisselev:Giudice:05}-\cite{Kisselev:Kisselev:06}.

The classical action of the model is
\begin{eqnarray}\label{Kisselev:action}
S &=& \int \!\! d^4x \!\! \int_{-\pi r_c}^{\pi r_c} \!\! dy \,
\sqrt{G} \, (2 \bar{M}_5^3 \mathcal{R}
- \Lambda) \nonumber \\
&+& \int \!\! d^4x \sqrt{|g^{(1)}|} \, (\mathcal{L}_1 - \Lambda_1) +
\int \!\! d^4x \sqrt{|g^{(2)}|} \, (\mathcal{L}_2 - \Lambda_2) \;,
\end{eqnarray}
where $G_{MN}(x,y)$ is the 5-dimensional metric, with $M,N =
0,1,2,3,4$, $\mu = 0,1,2,3$, and $y$ is the 5-th dimension
coordinate of the size $r_c$. The quantities
\begin{equation}
g^{(1)}_{\mu\nu}(x) = G_{\mu\nu}(x, y=0) \;, \quad
g^{(2)}_{\mu\nu}(x) = G_{\mu\nu}(x, y=\pi r_c)
\end{equation}
are induced metrics on the branes, $\mathcal{L}_1$ and
$\mathcal{L}_2$ are brane Lagrangians, $G = \det(G_{MN})$, $g^{(i)}
= \det(g^{(i)}_{\mu\nu})$.

As in the original RS model~\cite{Kisselev:Randall:99}, the
periodicity $y = y + 2\pi r_c$ is imposed and the points $(x_{\mu},
y)$ and $(x_{\mu}, -y)$ are identified. Thus, we get the orbifold
$S^1/Z_2$. There are two 3D branes located at the fixed points $y =
0$ (Plank brane) and $y = \pi r_c$ (TeV brane). The SM fields are
constrained to the TeV brane, while the gravity propagates in all
spatial dimensions.

In order to solve Einstein-Hilbert's equations which follow from the
action (\ref{Kisselev:action}), it is assumed that the background
metric respects 4-dimensional Poincare invariance
\begin{equation}\label{Kisselev:RS_background_metric}
\quad ds^2 = e^{-2 \sigma (y)} \, \eta_{\mu \nu} \, dx^{\mu} \,
dx^{\nu} - dy^2 \;,
\end{equation}
Thus, the 5-dimensional metric tensor has the form
\begin{equation}\label{cov_metric_tensor}
G_{M\!N} = \left(
  \begin{array}{cc}
  g_{\mu\nu} & 0 \\
    0 & -1 \\
  \end{array}
\right) \;,
\end{equation}
where
\begin{equation}\label{RS_4dim_metric}
g_{\mu\nu} = e^{-2 \sigma (y)} \, \eta_{\mu\nu} \;,
\end{equation}
and $\eta_{\mu\nu}$ is the Minkowski tensor $(1,-1,-1,-1)$.

The background metric was found to be
\cite{Kisselev:Kisselev:dielectron}
\begin{equation}\label{Kisselev:sigma}
\sigma(y) = \frac{\kappa}{2} \, ( |y| - |\pi r_c - y|) -
\frac{\kappa \pi r_c }{2} \;,
\end{equation}
with the fine tuning
\begin{eqnarray}\label{Kisselev:fine_tuning}
\Lambda &=& -6 \bar{M}_5^3\kappa^2 [ \varepsilon(y) +
\varepsilon(\pi r_c -
y)]^2 \;, \nonumber \\
\Lambda_1  &=&  - \Lambda_2 = 12 \bar{M}_5^3 \kappa \;.
\end{eqnarray}
The quantity $\kappa$ defines the curvature of the 5-dimensional
space-time.

In the RSSC model the hierarchy relation looks like
\begin{equation}\label{Kisselev:hierarchy_relation}
\bar{M}_{\mathrm{Pl}}^2 = \frac{\bar{M}_5^3}{\kappa} \left(e^{2 \pi
\kappa r_c} - 1 \right) \;,
\end{equation}
where $\bar{M}_{\mathrm{Pl}} = M_{\mathrm{Pl}} /\sqrt{8\pi}$ is the
reduced Planck scale, and $\bar{M}_5$ is the \emph{reduced}
fundamental gravity scale. $\bar{M}_5$ is related to the
5-dimensional Planck scale as
\begin{equation}\label{Kisselev:grav_scale_reduced}
\bar{M}_5 = (2\pi)^{-1/3} M_5 \;.
\end{equation}

The masses of the Kaluza-Klein (KK) graviton excitations
$h_{\mu\nu}^{(n)}$ are proportional to $\kappa$,
\begin{equation}\label{Kisselev:graviton_masses}
m_n = x_n \kappa \;, \quad n=1,2, \ldots \;.
\end{equation}
The interaction Lagrangian of gravitons is given by
\cite{Kisselev:Giudice:05}, \cite{Kisselev:Kisselev:05}
\begin{eqnarray}\label{Kisselev:Lagrangian_TeV}
\mathcal{L}_{\mathrm{int}} &=& - \frac{1}{\bar{M}_5^{3/2}}
\sum_{n=1}^{\infty} \int \!\! dy \, \sqrt{G} \,
h_{\mu\nu}^{(n)}(x,y) \, T_{\alpha\beta}(x) \, g^{\mu\alpha}
g^{\nu\beta} \delta (y - \pi \kappa r_c) \nonumber \\
&=& - \frac{1}{\Lambda_{\pi}} \, T_{\alpha\beta} (x)
\sum_{n=1}^{\infty} h^{(n)}_{\mu \nu} (x) \, \eta^{\mu\alpha}
\eta^{\nu\beta}\;,
\end{eqnarray}
where $T_{\alpha\beta}(x)$ is the energy-momentum tensor of the SM
fields on the TeV brane, and
\begin{equation}\label{Kisselev:lambda}
\Lambda_{\pi} = \bar{M}_{\mathrm{Pl}} \, e^{-\pi \kappa r_c} \;.
\end{equation}

Note that in the RSSC model the mass spectrum
(\ref{Kisselev:graviton_masses}) and experimental signature are
similar to those in the ADD model with one flat extra
dimension~\cite{Kisselev:Arkani-Hamed:98}. Let us remember that the
original RS model predicts heavy graviton resonances.

\section{Graviton contribution to dilepton production at the LHC}

In the framework of the RSSC model~\cite{Kisselev:Kisselev:05},
\cite{Kisselev:Kisselev:06}, gravity effects can be searched for in
the dilepton production ($l = \mu \mathrm{\ or \ } e$),
\begin{equation}\label{Kisselev:process}
p \, p \rightarrow l^+ l^- + X \;.
\end{equation}
In particular, the $p_{\perp}$-distribution of the final leptons in
the process (\ref{Kisselev:process}) is given by the formula
\begin{eqnarray}\label{Kisselev:cross_sec}
\frac{d \sigma}{d p_{\perp}}(p p \rightarrow l^+ l^- + X) &=&
2p_{\perp} \!\!\!\! \sum\limits_{a,b = q,\bar{q},g} \!\!
\int\nolimits \!\! \frac{d\tau \sqrt{\tau}}{\sqrt{\tau -
x_{\perp}^2}} \! \int\nolimits \! \frac{dx_1}{x_1}  f_{a/p}(\mu^2,
x_1)
\nonumber \\
&\times& f_{b/p}(\mu, \tau/x_1) \, \frac{d \hat{\sigma}}{d\hat{t}}(a
b \rightarrow  l^+ l^-) \;,
\end{eqnarray}
Here $f_{c/p}(\mu^2, x)$ is the distribution of the parton of the
type $c$ in momentum fraction $x$ inside the proton taken at the
scale $\mu$. $d\hat{\sigma}/d\hat{t}$ is the differential cross
section of the subprocess $a b \rightarrow l^+ l^-$. In
eq.~(\ref{Kisselev:cross_sec}) two dimensionless variables are
introduced
\begin{equation}\label{Kisselev:tau_xtr}
x_{\perp} = \frac{2 p_{\perp}}{\sqrt{s}} \;, \quad \tau = x_1 x_2
\,,
\end{equation}
where $x_2$ is the momentum fraction of the parton $b$ inside the
proton.

The contribution of the virtual gravitons to lepton pair production
comes from the quark-antiquark annihilation and gluon-gluon fusion,
\begin{eqnarray} \label{Kisselev:parton_cross_sec}
\frac{d\hat{\sigma}}{d\hat{t}}(q \bar{q} \rightarrow l^+l^-) &=&
\frac{\hat{s}^4 + 10\hat{s}^3 \hat{t} + 42 \, \hat{s}^2 \hat{t}^2 +
64 \hat{s} \, \hat{t}^3 + 32 \, \hat{t}^4}{1536 \, \pi \hat{s}^2}
\left| \mathcal{S}(\hat{s}) \right|^2 \;,
\nonumber \\
\frac{d\hat{\sigma}}{d\hat{t}}(gg \rightarrow l^+l^-) &=&
-\frac{\hat{t}(\hat{s} + \hat{t}) (\hat{s}^2 + 2 \hat{s}\,\hat{t} +
2\,\hat{t}^2)}{256 \, \pi \hat{s}^2} \left|\mathcal{S}(\hat{s})
\right|^2 \;,
\end{eqnarray}
where $\hat{s}$ and $\hat{t}$ are Mandelstam variables of the
subprocess. The sum
\begin{equation}\label{Kisselev:KK_summation}
\mathcal{S}(s) = \frac{1}{\Lambda_{\pi}^2} \sum_{n=1}^{\infty}
\frac{1}{s - m_n^2 + i \, m_n \Gamma_n} \;
\end{equation}
is the invariant part of the partonic matrix elements, with
$\Gamma_n$ being total width of the graviton with the KK number $n$
and mass $m_n$~\cite{Kisselev:Kisselev:06}
\begin{equation}\label{Kisselev:graviton_widths}
\Gamma_n = \eta \, m_n \left( \frac{m_n}{\Lambda_{\pi}} \right)^2,
\quad \eta \simeq 0.09 \;.
\end{equation}
Note that the function $\mathcal{S}(s)$ is universal for all
processes mediated by $s$-channel virtual gravitons.

In the RSSC model \cite{Kisselev:Kisselev:06} the explicit
expression was obtained for $\mathcal{S}(s)$
(\ref{Kisselev:KK_summation}) at $s \sim \bar{M}_5 \gg \kappa$

\begin{equation}\label{Kisselev:KK_sum}
\mathcal{S}(s) = - \frac{1}{4 \bar{M}_5^3 \sqrt{s}} \; \frac{\sin 2A
+ i \sinh 2\varepsilon }{\cos^2 \! A + \sinh^2 \! \varepsilon} \;,
\end{equation}
where
\begin{equation}\label{Kisselev:parameters}
A = \frac{\sqrt{s}}{\kappa} \;, \qquad \varepsilon = \frac{\eta}{2}
\Big( \frac{\sqrt{s}}{\bar{M}_5} \Big)^3 \;.
\end{equation}

In papers \cite{Kisselev:Kisselev:dimuon},
\cite{Kisselev:Kisselev:dielectron} contributions from $s$-channel
gravitons to the $p_{\perp}$-distributions of the final leptons were
calculated in the RSSC model by using
eqs.~(\ref{Kisselev:cross_sec})- (\ref{Kisselev:parameters}). The
calculations were made for different values of 5-dimensional Planck
scale $\bar{M}_5$. The MSTW 2008 NNLO parton
distributions~\cite{Kiselev:MSTW} were used. The PDF scale $\mu$ was
taken to be equal to the invariant mass of the lepton pair, $\mu =
M_{l^+l^-} = \sqrt{\hat{s}}$.

The CMS cuts  on the lepton pseudorapidities were imposed. For the
dimuon events the cut looks like
\begin{equation}\label{Kisselev:rapidity_cut_mu}
|\eta| < 2.4 \;,
\end{equation}
while for the dielectron events the cuts are the following
\begin{equation}\label{Kisselev:rapidity_cut_el}
|\eta| < 1.44 \;, \quad  1.57 < |\eta| < 2.50 \;.
\end{equation}
The reconstruction efficiency of 85$\%$ was assumed for the dilepton
events \cite{Kisselev:CMS_dilepton_efficiency}.

\begin{figure}
\begin{minipage}{.45\textwidth}
\centering
\includegraphics[bb= 77 273 483 540, width=.9\textwidth]{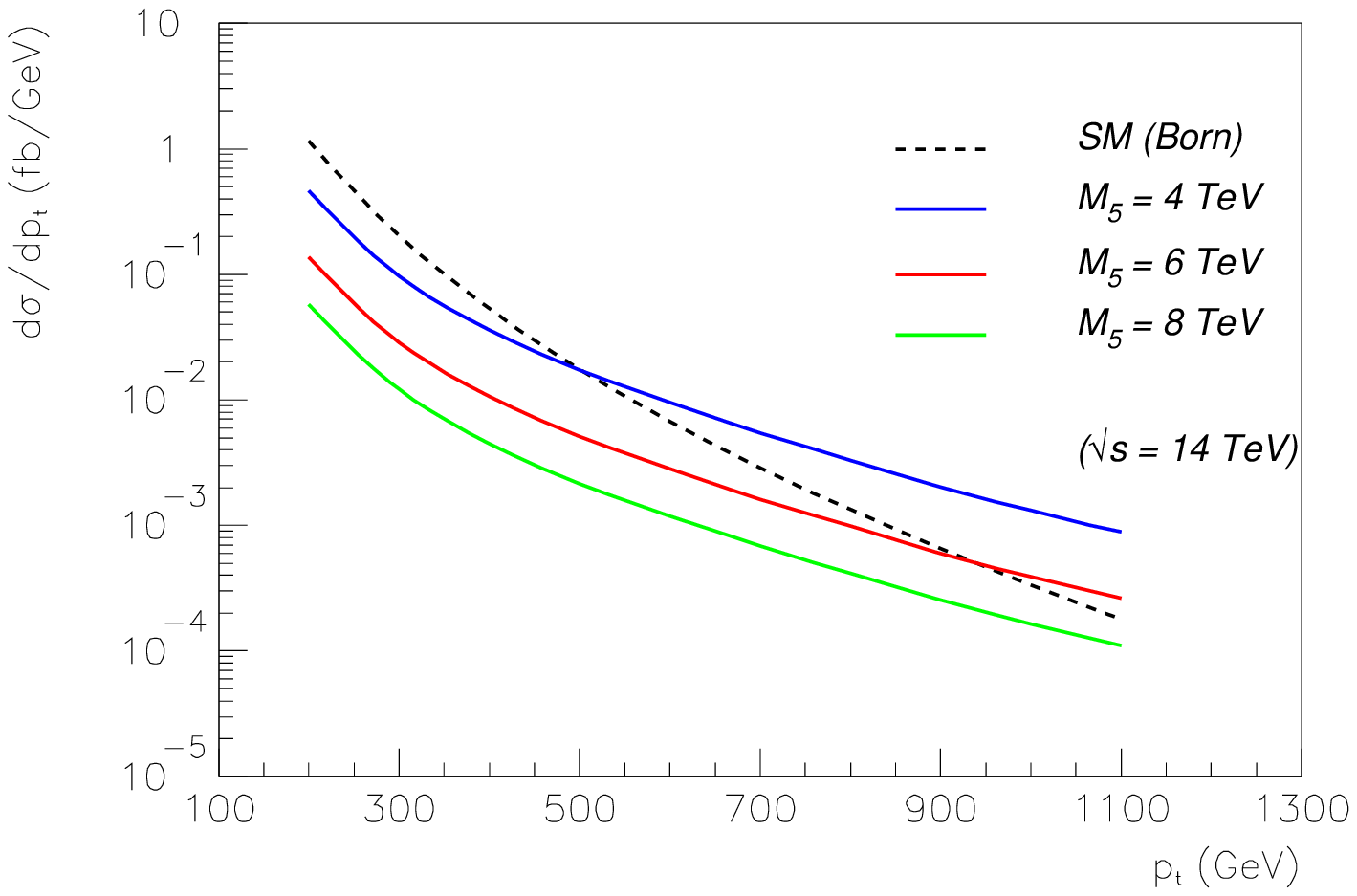}
\caption{The KK graviton contribution to the dimuon production at
the LHC for several values of 5-dimensional reduced Planck scale
(solid curves) vs. SM contribution (dashed curve) at $\sqrt{s} = 14$
TeV.} \label{Kisselev:fig:1}
\end{minipage}
 \rule{.05\textwidth}{0pt}
\begin{minipage}{.45\textwidth}
\centering
\includegraphics[bb= 77 273 483 540,width=.9\textwidth]{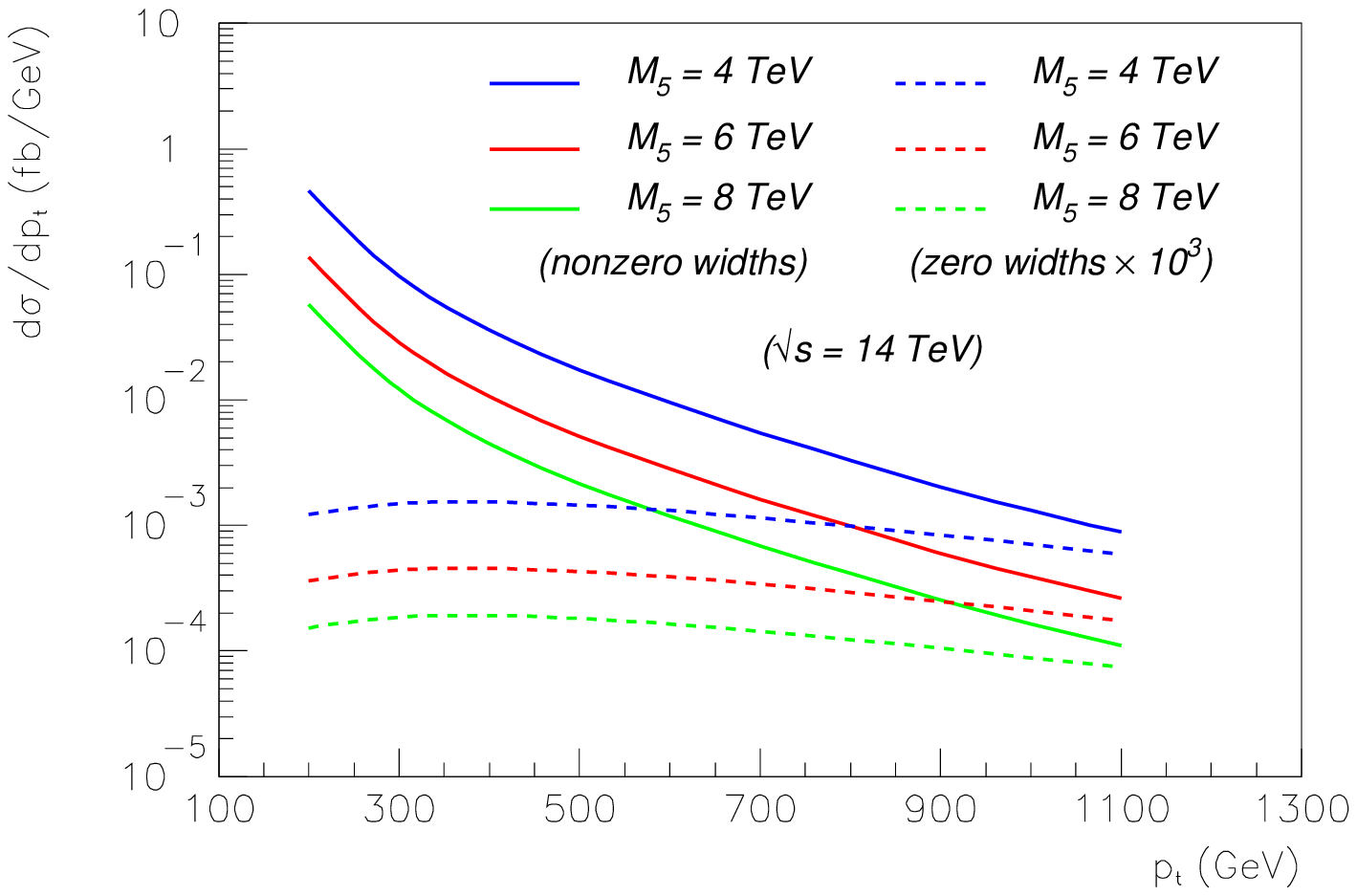}
\caption{The contributions to the dimuon events from the KK
gravitons with nonzero widths (solid curves) vs. contributions from
zero width gravitons (multiplied by 10$^3$, dashed curves).}
\label{Kisselev:fig:2}
\end{minipage}
\end{figure}

\begin{figure}
\begin{minipage}[t]{.45\textwidth}
\centering
\includegraphics[bb= 0 0 360 346, width=.9\textwidth]{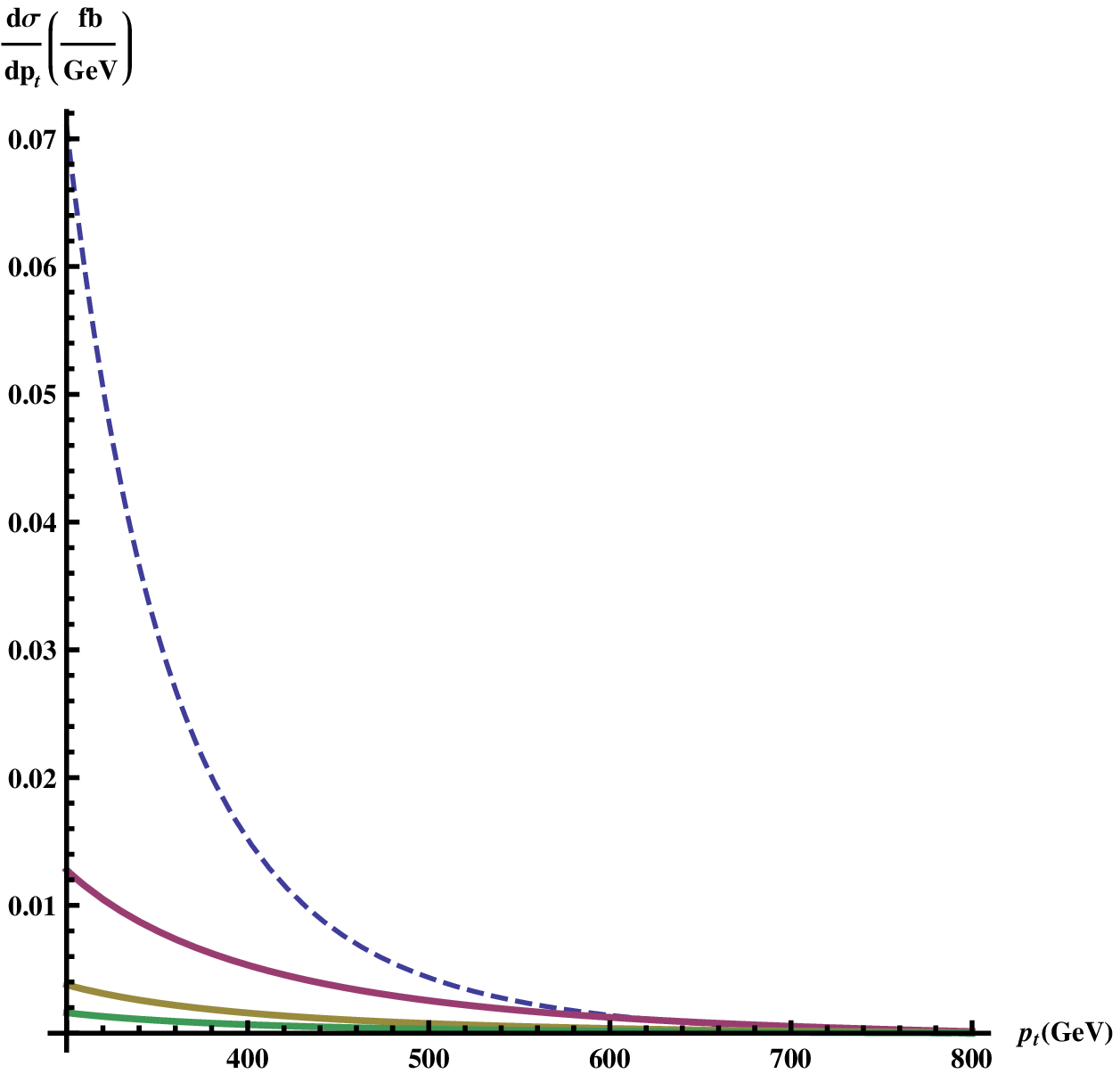}
\caption{The KK graviton contribution to the dielectron production
for $M_5$ = 2, 4, 6 TeV (solid curves, from above) vs. SM (Born)
contribution (dashed curve) at $\sqrt{s} = 8$ TeV.}
\label{Kisselev:fig:3}
\end{minipage}
 \rule{.05\textwidth}{0pt}
\begin{minipage}[t]{.45\textwidth}
\centering
\includegraphics[bb= 0 0 360 346 ,width=.9\textwidth]{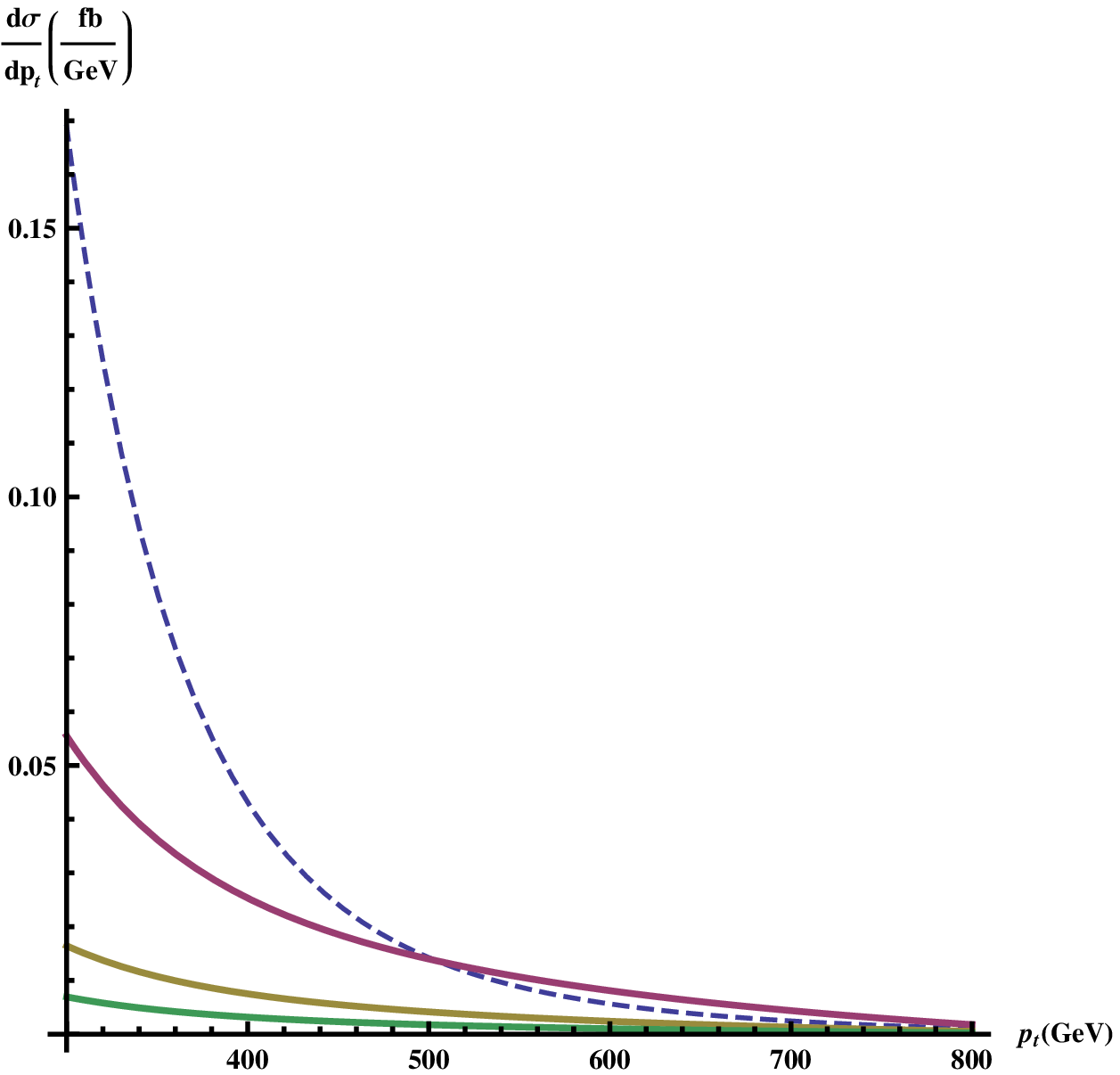}
\caption{The same as in Fig.~\ref{Kisselev:fig:4}, but for $M_5$ =
4, 6, 8 TeV and $\sqrt{s} = 13$ TeV.}
\label{Kisselev:fig:4}
\end{minipage}
\end{figure}

In Fig.~\ref{Kisselev:fig:1} the gravity cross sections for the
dimuon production at the LHC are presented. The gravity mediated
contributions to the cross sections do not include the SM
contribution. Fig.~\ref{Kisselev:fig:2} demonstrates that an
ignorance of the graviton widths would be a rough approximation
since it results in very large suppression of the gravity
contributions. The $p_{\perp}$-distributions for the dielectron
production are shown in Fig.~\ref{Kisselev:fig:3} and
Fig.~\ref{Kisselev:fig:4}, for $\sqrt{s} = 8$ TeV and $\sqrt{s} =
13$ TeV, respectively.

\begin{figure}
\begin{minipage}[t]{.45\textwidth}
\centering
\includegraphics[bb= 0 0 360 319, width=.9\textwidth]{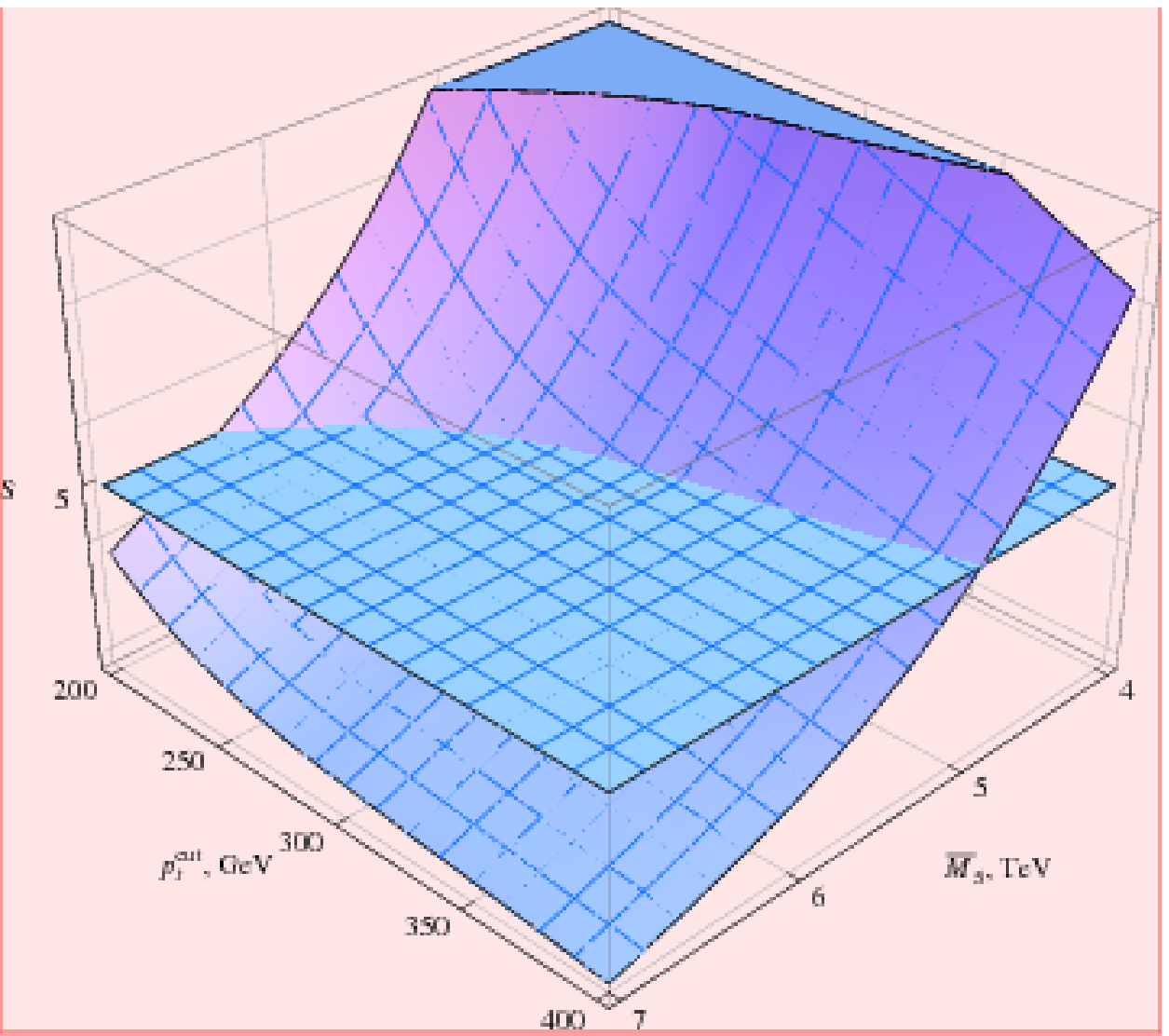}
\caption{The statistical significance $S$ for the dimuon production
at the LHC for $\sqrt{s} = 7$ TeV and integrated luminosity 5
fb$^{-1}$ as a function of the transverse momentum cut
$p_{\perp}^{\mathrm{cut}}$ and reduced gravity scale $\bar{M}_5$.
The plane $S=5$ is also shown.} \label{Kisselev:fig:5}
\end{minipage}
 \rule{.05\textwidth}{0pt}
\begin{minipage}[t]{.45\textwidth}
\centering
\includegraphics[bb= 0 0 360 318, width=.9\textwidth]{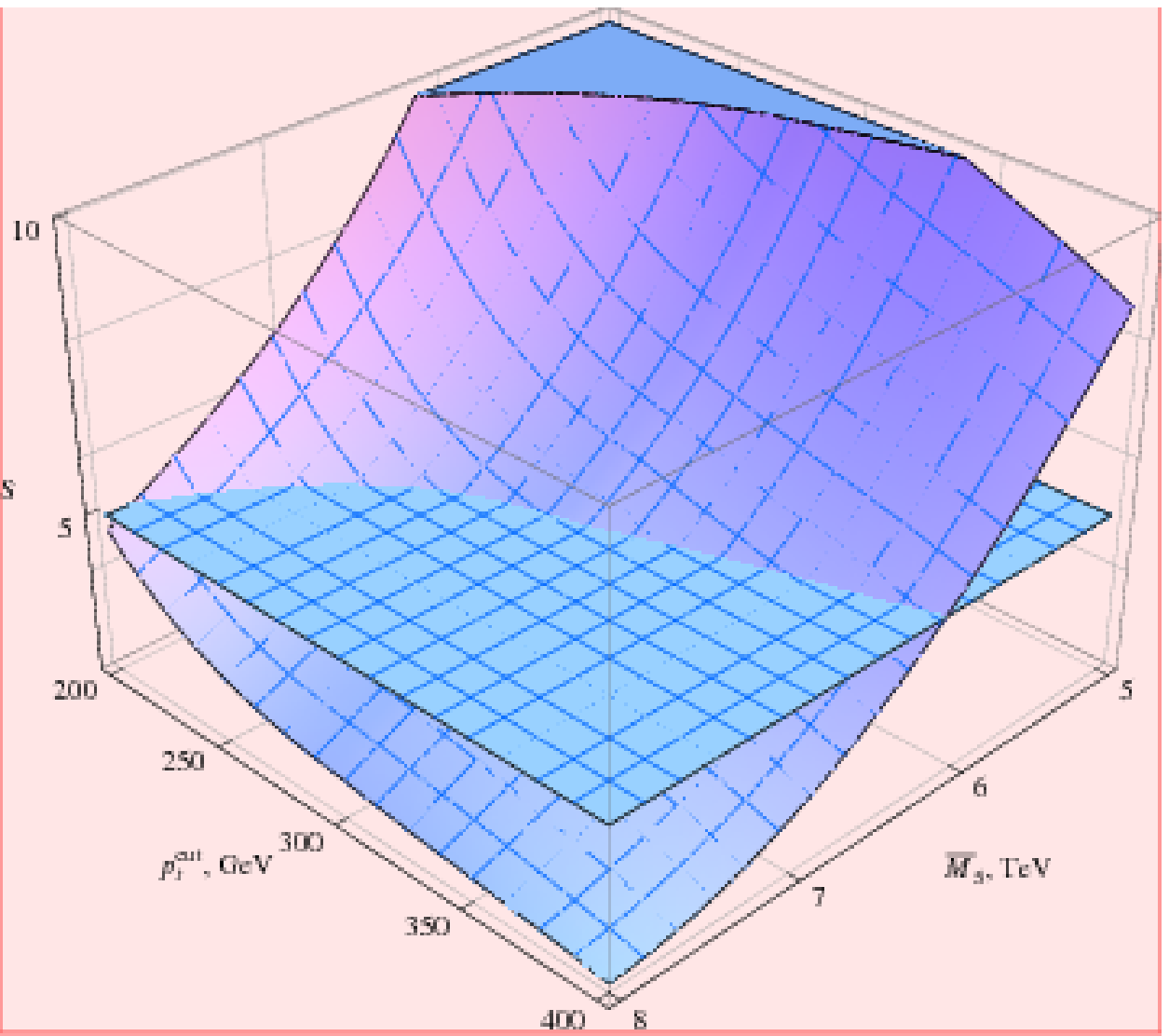}
\caption{The same as in Fig.~\ref{Kisselev:fig:5}, but for $\sqrt{s}
= 14$ TeV and integrated luminosity 30 fb$^{-1}$.}
\label{Kisselev:fig:6}
\end{minipage}
\end{figure}

\begin{figure}
\begin{minipage}[t]{.45\textwidth}
\centering
\includegraphics[bb= 0 0 360 319, width=.9\textwidth]{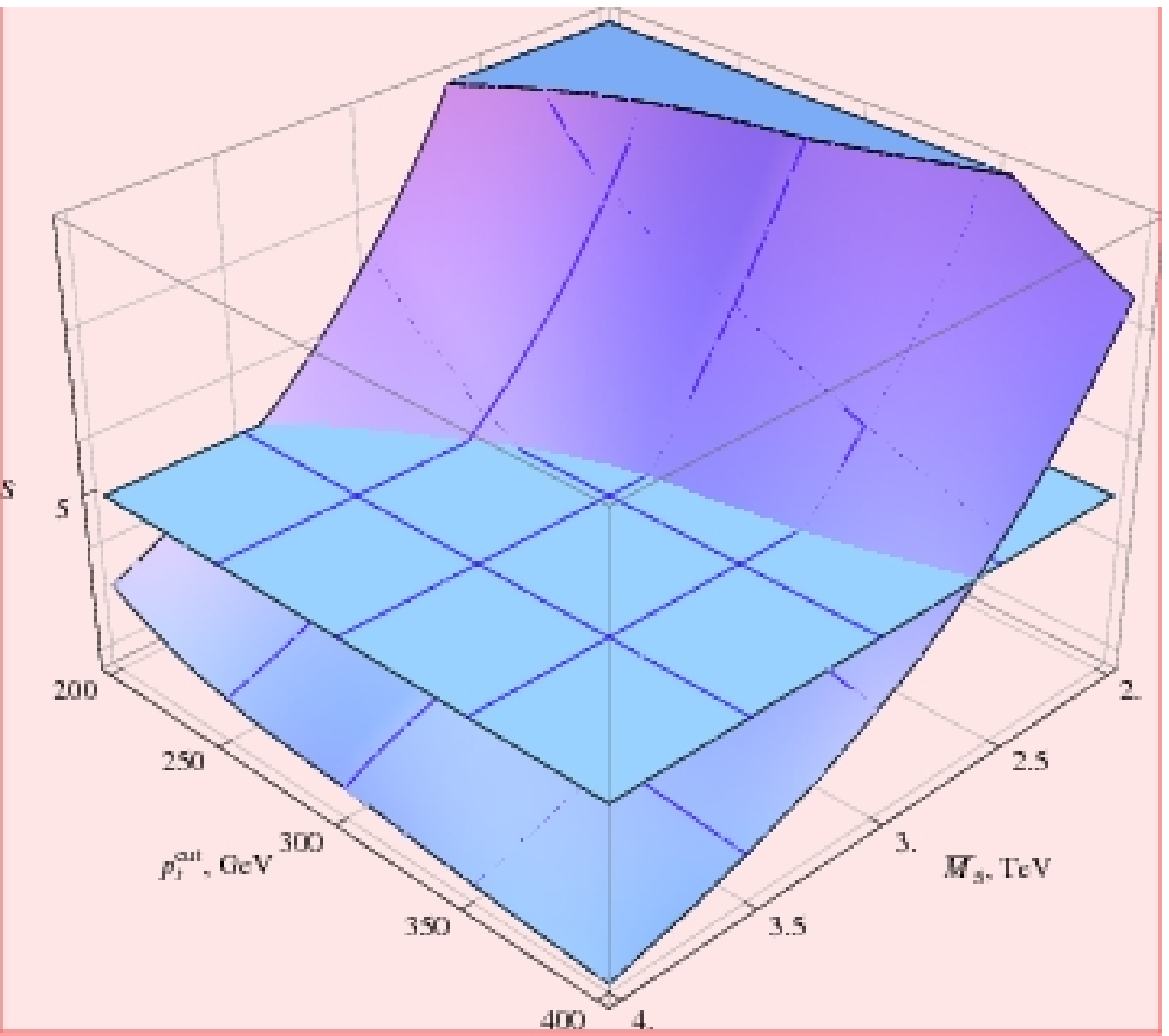}
\caption{The statistical significance for the dielectron
 production at the LHC for $\sqrt{s} = (7+8)$ TeV and integrated
 luminosity (5+20) fb$^{-1}$.}
 \label{Kisselev:fig:7}
\end{minipage}
 \rule{.05\textwidth}{0pt}
\begin{minipage}[t]{.45\textwidth}
\centering
\includegraphics[bb= 0 0 360 319, width=.9\textwidth]{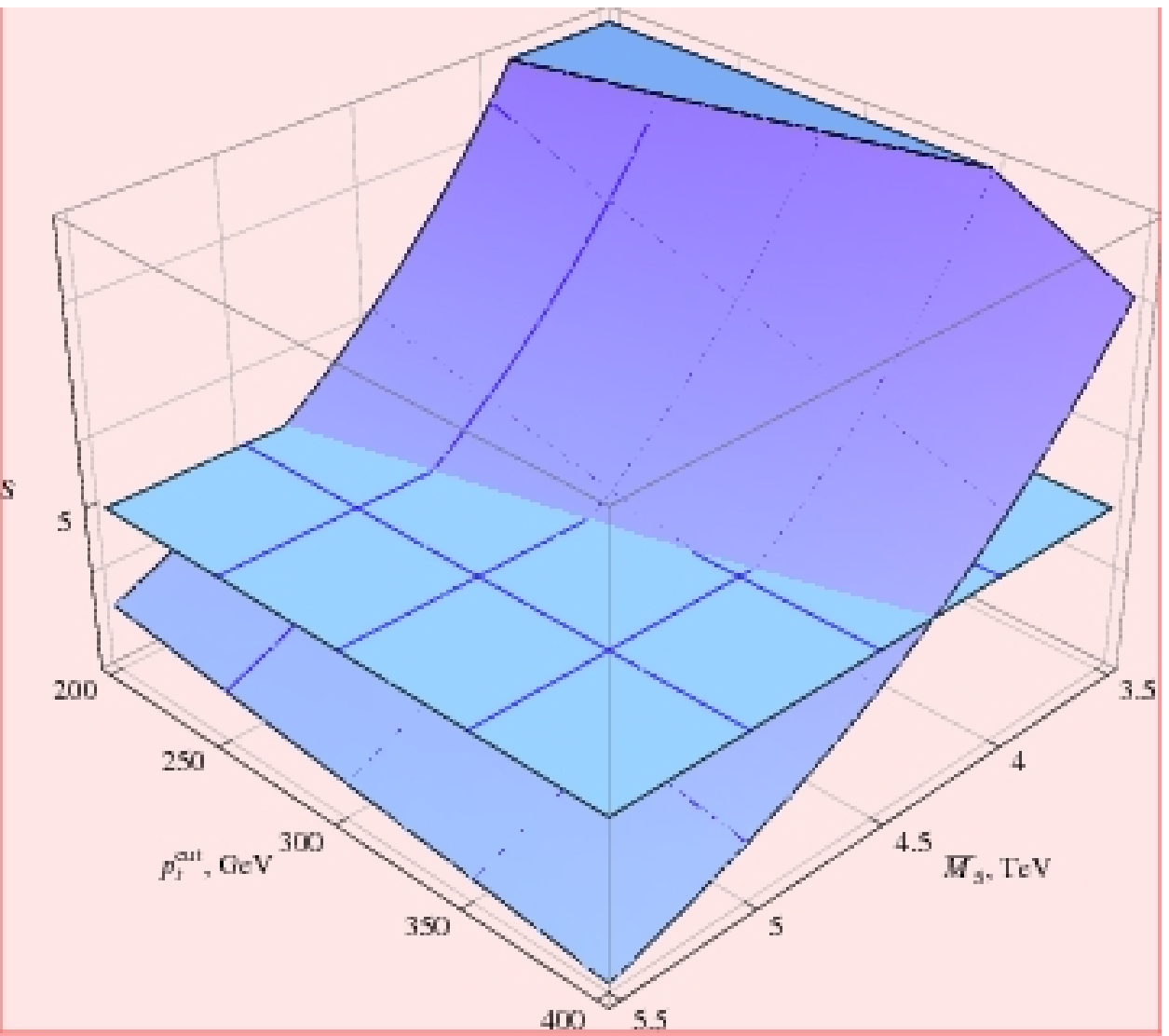}
\caption{The same as in Fig.~\ref{Kisselev:fig:7}, but for
 $\sqrt{s} = 13$ TeV and integrated luminosity 30 fb$^{-1}$.}
 \label{Kisselev:fig:8}
\end{minipage}
\end{figure}

Let $N_S$($N_B$) be a number of signal (background) dilepton events
with $p_{\perp} > p_{\perp}^{\mathrm{cut}}$,
\begin{equation}\label{Kisselev:ev_numder}
N_B = \!\! \int\limits_{p_{\perp} > p_{\perp}^{\mathrm{cut}}} \!\!
\frac{d \sigma (\mathrm{SM})}{dp_{\perp}} dp_{\perp} \;, \quad N_S =
\!\! \int\limits_{p_{\perp} > p_{\perp}^{\mathrm{cut}}} \!\! \frac{d
\sigma (\mathrm{grav})}{dp_{\perp}} dp_{\perp} \;.
\end{equation}
Then one can define the statistical significance
\begin{equation}\label{Kisselev:significance}
\mathcal{S} = \frac{N_S}{\sqrt{N_B + N_S}} \;,
\end{equation}
and require a $5 \sigma$ effect. In Fig.~\ref{Kisselev:fig:5} the
statistical significance is shown for $\sqrt{s} = 7$ TeV as a
function of the transverse momentum cut $p_{\perp}^{\mathrm{cut}}$
and reduced 5-dimensional Planck scale $\bar{M}_5$.
Fig.~\ref{Kisselev:fig:6} represents the statistical significance
for the dimuon events at $\sqrt{s} = 14$ TeV. The statistical
significances for the dielectron events are shown in
Fig.~\ref{Kisselev:fig:7} and Fig.~\ref{Kisselev:fig:8}.

To take into account higher order contributions, the $K$-factor of
1.5 for the SM background was taken, while the factor $K=1$ was used
for the signal.

As a result, LHC discovery limits on the 5-dimensional Planck scale
$M_5$ were obtained. In particular, for the (7+8) TeV LHC with the
integrated luminosity (5+20) fb$^{-1}$ the search limit in the
dielectron production is equal
to~\cite{Kisselev:Kisselev:dielectron}
\begin{equation}\label{Kisselev:search limit_7+8_TeV}
M_5 = 6.35 \mathrm{\ TeV} \;.
\end{equation}
For the 13 TeV dielectron events with the integrated luminosity 30
fb$^{-1}$ the search limit was found to be
\cite{Kisselev:Kisselev:dielectron}
\begin{equation}\label{Kisselev:search limit_13_TeV}
M_5 = 8.95 \mathrm{\ TeV} \;.
\end{equation}

Let us underline that in the RSSC model these bounds do not depend
on the curvature $\kappa$, contrary to the original RS
model~\cite{Kisselev:Randall:99}.



\end{document}